\documentclass[aps, superscriptaddress]{revtex4-1}

\pdfoutput=1
\usepackage{graphicx}

\usepackage{amsmath}
\usepackage{amssymb}
\usepackage{bm}
\usepackage{color}

\newcommand{\bra}[1]{\langle{#1}|}
\newcommand{\ket}[1]{|{#1}\rangle}

\newcommand{\Tr}{{\rm Tr}\hspace{0.07cm}}

\newcommand{\abs}[1]{{|#1|}}

\begin{document}
	\title{Instantaneous phase synchronization of two decoupled quantum limit-cycle
		oscillators induced by conditional photon detection}
	\author{Yuzuru Kato}
	\email{Corresponding author: kato.y.bg@m.titech.ac.jp}
	\affiliation{Department of Systems and Control Engineering,
		Tokyo Institute of Technology, Tokyo 152-8552, Japan}
	
	\author{Hiroya Nakao}
	\affiliation{Department of Systems and Control Engineering,
		Tokyo Institute of Technology, Tokyo 152-8552, Japan}
	\date{\today}
	
	\begin{abstract}
		We show that conditional photon detection induces instantaneous phase
		synchronization between two decoupled quantum limit-cycle oscillators.
		We consider two quantum van der Pol oscillators without mutual coupling, each
		with additional single-photon dissipation, and perform continuous measurement of
		photon counting on the output fields of the 
		two baths 
		interacting through a beam splitter.
		It is observed that in-phase or anti-phase coherence 
		of the two decoupled oscillators instantaneously increases after the
		photon detection 
		and then decreases gradually in the weak quantum regime
		or quickly in the strong quantum regime until the next photon detection occurs.
		In the strong quantum regime, quantum entanglement also increases after the
		photon detection and quickly disappears.
		We derive the analytical upper bounds for the increases
		in the quantum entanglement and phase coherence
		by the conditional photon detection in the quantum limit. 
	\end{abstract}
	
	\maketitle
	
	
	\section{Introduction}
	%
	Synchronization phenomena, first reported by Huygens in the 17th century, are
	widely observed in various areas of science and engineering,
	including laser oscillations, 
	mechanical vibrations, oscillatory chemical reactions, and biological
	rhythms~\cite{winfree2001geometry, kuramoto1984chemical, 
		pikovsky2001synchronization, nakao2016phase, 
		ermentrout2010mathematical, strogatz1994nonlinear}.
	While synchronization of \textit{coupled} 
	or \textit{periodically driven} nonlinear oscillators has been extensively
	investigated \cite{winfree2001geometry, kuramoto1984chemical, 
		pikovsky2001synchronization, aronson1990amplitude}, 
	decoupled oscillators
	that \textit{do not involve any interactions or periodic forcing} 
	can also
	exhibit synchronous behaviors when driven by common random forcing, 
	such as 
	consistency or reproducibility of laser oscillations
	and spiking neocortical neurons receiving identical sequences of random
	signals~\cite{uchida2004consistency,mainen1995reliability}.
	The common-noise-induced synchronization has been theoretically investigated for
	decoupled limit-cycle oscillators subjected, e.g., to common random
	impulses~\cite{pikovskii1984synchronization,nakao2005synchrony,arai2008phase}
	and Gaussian white noise~\cite{teramae2004robustness,
		goldobin2005synchronization, nakao2007noise}.

	Recent developments in nanotechnology have inspired theoretical
	investigations of quantum synchronization \cite{
			lee2013quantum,
			lee2014entanglement,
			walter2015quantum,
			xu2014synchronization,
			roulet2018quantum,
			mari2013measures,ameri2015mutual, galve2017quantum,
			lorch2017quantum, nigg2018observing,
			xu2015conditional,
			weiss2016noise,
			es2020synchronization,chia2020relaxation, kato2020enhancement,
			koppenhofer2019optimal,
			walter2014quantum, sonar2018squeezing,
			lorch2016genuine,  weiss2017quantum,
			roulet2018synchronizing,  
			witthaut2017classical,
			kato2019semiclassical, kato2020semiclassical, kato2020quantum,
			mok2020synchronization},
	and the first experimental demonstration of 
	quantum phase synchronization in spin-1 atoms~\cite{laskar2020observation} and
	on the
	IBM Q system~\cite{koppenhofer2020quantum} has been reported very recently.
	Many studies have analyzed
	coupled quantum nonlinear dissipative oscillators, 
	for example,
	synchronization of quantum van der Pol (vdP)
	oscillators~\cite{lee2013quantum,lee2014entanglement, walter2015quantum}, 
	synchronization of ensembles of atoms~\cite{xu2014synchronization}, 
	synchronization of triplet spins~\cite{roulet2018quantum},
	measures for quantum synchronization of two oscillators~\cite{mari2013measures,
		ameri2015mutual, galve2017quantum}, 
	and synchronization blockade~\cite{lorch2017quantum, nigg2018observing}.
	The effects of quantum measurement backaction 
	on quantum nonlinear dissipative oscillators
	have also been investigated as a unique feature of quantum systems,
	including improvement in the accuracy of Ramsey spectroscopy through measurement
	of synchronized atoms~\cite{xu2015conditional}, 
	measurement-induced transition between in-phase and anti-phase synchronized
	states~\cite{weiss2016noise}, 
	unraveling of nonclassicality in optomechanical
	oscillators~\cite{koppenhofer2018unraveling}, 
	characterization of synchronization using quantum
	trajectories~\cite{es2020synchronization},
	realization of quantum relaxation oscillators
		\cite{chia2020relaxation},
	and enhancement of synchronization by quantum measurement 
	and feedback control~\cite{kato2020enhancement}.
	
	In this study, inspired by the common-noise-induced synchronization of decoupled
	classical oscillators,
	we consider phase synchronization of two decoupled quantum oscillators
	induced by common backaction of quantum measurement.
	We consider two quantum van der Pol oscillators without mutual coupling, each
	with 
	additional single-photon dissipation, and perform continuous
	measurement of photon counting on the output fields of the 
	two baths 
	interacting through a beam splitter.
	It is demonstrated that the quantum measurement backaction of conditional photon
	detection common to both oscillators induces instantaneous phase synchronization
	of the oscillators.

	\begin{figure} [!t]
		\begin{center}
			\includegraphics[width=0.9\hsize,clip]{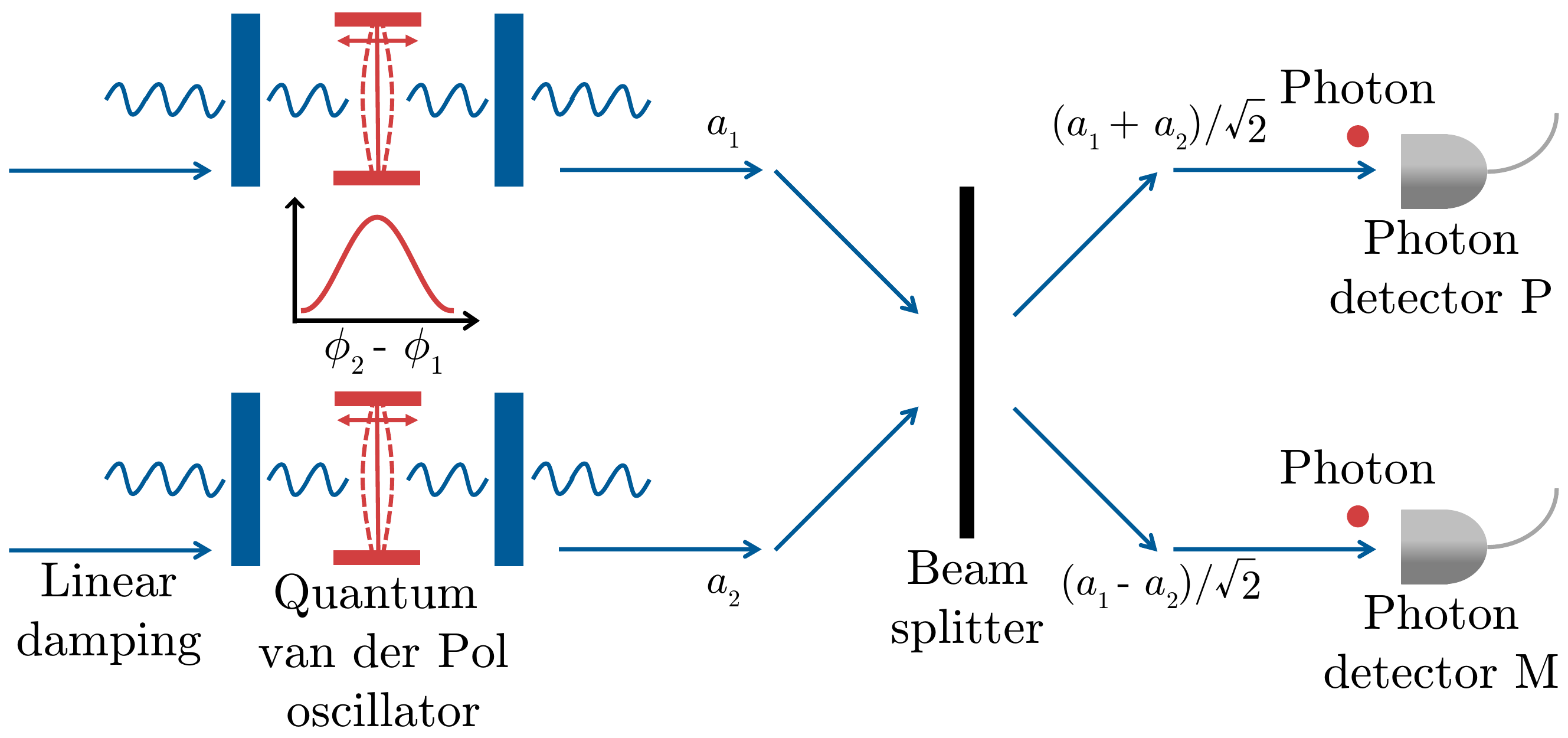}
			\caption{
				Instantaneous phase synchronization of 
				two decoupled quantum vdP oscillators induced by 
				conditional photon detection.
				Either in-phase or anti-phase coherence is induced
				after photon detection at detector P or M, respectively.
			}
			\label{fig_1}
		\end{center}
	\end{figure}
	%

	
	\section{Model}
	A schematic of the physical setup is depicted in Fig.~\ref{fig_1}.
	The stochastic master equation (SME) of the system can be
	expressed as 
	\begin{align}
		\label{eq:qvdp_sme}
		d \rho &= \mathcal{L}_0\rho dt
		+\mathcal{G}[L_{+}]\rho 
		\left( dN_{+} - \gamma_{3} \Tr[ L_{+}^{\dag} L_{+} \rho] dt \right)
		+\mathcal{G}[L_{-}]\rho 
		\left( dN_{-} - \gamma_{3} \Tr[ L_{-}^{\dag} L_{-} \rho] dt \right),
		\cr
		\mathcal{L}_0 \rho
		&=
		\sum_{j=1,2} \left( -i \left[\omega a_{j}^{\dag}a_{j}, \rho \right]
		+ \gamma_{1} \mathcal{D}[a_{j}^{\dag}]\rho 
		+ \gamma_{2}\mathcal{D}[a_{j}^{2}]\rho 
		+ \gamma_{3}\mathcal{D}[a_{j}]\rho 
		\right),
		\cr
		L_{\pm} &= \frac{1}{\sqrt{2}}(a_1 \pm a_2),~
		\mathcal{D}[L]\rho = L \rho L^{\dag} - 
		\frac{1}{2}\left(\rho L^{\dag} L + L^{\dag} L \rho \right),~
		\mathcal{G}[L]\rho = \frac{L \rho L^{\dag}}{ \Tr[L \rho L^{\dag}]} - \rho,
	\end{align}
	where the natural frequency $\omega$
	and the decay rates $\gamma_{1}$, $\gamma_{2}$, and $\gamma_{3}$ 
	for negative damping, nonlinear damping, and linear damping, respectively,
	are assumed identical for both oscillators,
	$N_{\pm}$ are two independent Poisson processes whose increments are given by
	$dN_{\pm} = 1$ with probability 
	$ \gamma_{3} \Tr[ L^{\dag}_{\pm}  L_{\pm} \rho ]dt$ and $dN_{\pm}
	= 0$ with probability 
		$1 - \gamma_{3} \Tr[ L^{\dag}_{\pm}  L_{\pm} \rho]dt$
	in each interval $dt$, where $dN_{+} = 1$ and $dN_{-} = 1$ represent the photon
	detection at detectors P and M
	in Fig.~\ref{fig_1}, respectively, and the reduced Planck constant is set to
	$\hbar = 1$.
	
	 In the derivation of the SME (\ref{eq:qvdp_sme}),
		the SLH framework~\cite{gough2009series, combes2017slh},
		a general formulation for quantum networked systems, has
		been used to describe the cascade and concatenate connections of the quantum
		system components.
		In this framework, the quantum system is specified by
		parameters $(\bm{S}, \bm{L}, H)$, i.e.,
		a scattering matrix $\bm S$, 
		coupling vector $\bm{L}$,
		and Hamiltonian $H$, from which 
		the SME (\ref{eq:qvdp_sme}) can be derived by using the quantum filtering theory
		\cite{van2005feedback,bouten2007introduction}.
		See Appendix~A for the details of the SLH framework and the derivation of the
		SME (\ref{eq:qvdp_sme}).

		Note that if we average Eq.~(\ref{eq:qvdp_sme}) over various stochastic
		trajectories,   we obtain two independent master equations without measurement
		for completely decoupled oscillators whose phase values are fully incoherent.
		The conditional photon detection at the detector P and M by the
		operators $L_+ = (a_1 + a_2)/\sqrt{2}$ and $L_- = (a_1 - a_2)/\sqrt{2}$
		after the beam splitter can also be interpreted as an unraveling of two competing dissipative coupling terms $D[a_1+a_2]$ and $D[a_1-a_2]$, which induce
		synchronization of the two oscillators \cite{lee2014entanglement,ishibashi2017oscillation}.
	
	
	%
	\section{Weak quantum regime}
	\begin{figure} [!t]
		\begin{center}
			\includegraphics[width=1\hsize,clip]{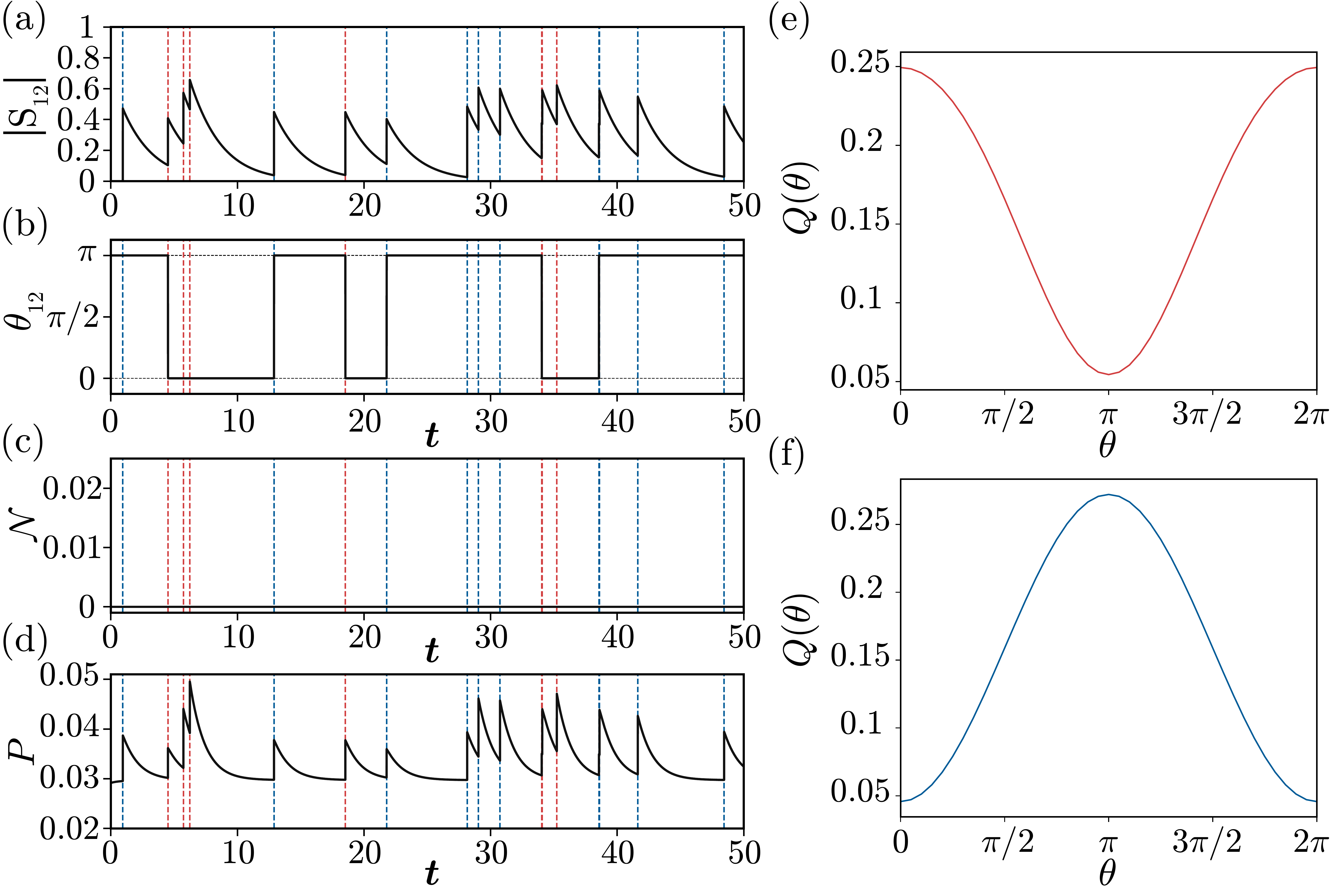}
			\caption{
				Results in the weak quantum regime. The parameters are $ (\omega, \gamma_2,
				\gamma_3)/\gamma_1 = (0.1, 0.25, 0.1)$ with $\gamma_1 = 1$.
				(a-d): Time evolution of 
				(a) absolute value of the normalized correlator $\abs{S_{12}}$,
				(b) average phase value $\theta_{12}$,
				(c) negativity $\mathcal{N}$,
				and
				(d) purity $P$.
				(e,f): Q distributions $Q(\theta)$ immediately after the first photon
				detection at (e) P ($t=4.51$) 
				and (f) M ($t=0.95$).			
				The photon detection at detectors P and M is indicated by the red- and
				blue-dashed lines in (a-d), respectively.
			}
			\label{fig_2}
		\end{center}
	\end{figure}
	First, we numerically analyze the quantum SME (\ref{eq:qvdp_sme}) in the weak
	quantum regime. 
	To characterize the degree of phase coherence between 
	the two quantum vdP oscillators, 
	we use the absolute value of the normalized correlator~\cite{weiss2016noise}
	\begin{align}
		\label{eq:norm_corr}
		S_{12} = \abs{S_{12}}e^{i \theta_{12}} = 
		\frac{ \Tr[ a_1^{\dag} a_2 \rho] }
		{\sqrt{ \Tr[ a_1^{\dag} a_1 \rho] \Tr[ a_2^{\dag} a_2 \rho]}}
	\end{align}
	as the order parameter,
	which is a quantum analog of the order parameter for two classical noisy
	oscillators~\cite{pikovsky2001synchronization}.
	The modulus $\abs{S_{12}}$ takes values in $0 \leq \abs{S_{12}} \leq 1$;
	$\abs{S_{12}} = 1$ when the two oscillators are
	perfectly phase-synchronized and $\abs{S_{12}} = 0$ when they are perfectly
	phase-incoherent.
	We also use the argument $\theta_{12}$ to characterize the averaged phase
	difference of the two oscillators
	in order to distinguish in-phase and anti-phase coherence.
	We use the negativity $\mathcal{N} =
	({\left\|\rho^{\Gamma_{1}}\right\|_{1}-1})/{2}$ to quantify the quantum
	entanglement of the two oscillators, where $\rho^{\Gamma_{1}}$ represents the
	partial transpose of the system with respect to the subsystem representing the
	first oscillator and 
	$\left\|X\right\|_{1}=\operatorname{Tr}|X|=\operatorname{Tr} \sqrt{X^{\dagger}
		X}$ \cite{zyczkowski1998volume, vidal2002computable}.
	When $\mathcal{N}$ takes a nonzero value, the two oscillators are entangled with one other.
	We also observe the purity $P = \Tr[\rho^2]$. 

	Figures~\ref{fig_2}(a),~\ref{fig_2}(b),~\ref{fig_2}(c), and \ref{fig_2}(d)
	plot the time evolution of $\abs{S_{12}}$, $\theta_{12}$, $\mathcal{N}$, and $P$
	in the weak quantum regime, respectively, calculated for a single trajectory of
	the quantum SME (\ref{eq:qvdp_sme}).
	As shown in Fig.~\ref{fig_2}(a), $\abs{S_{12}}$ instantaneously increases after
	the detection of a photon either at P or M,
	indicating that phase coherence of the two decoupled oscillators is induced by
	the conditional photon detection.
	After the photon detection, $\abs{S_{12}}$ gradually decreases because the two
	oscillators converge to the desynchronized steady state of the SME
	(\ref{eq:qvdp_sme}) in the absence of photon detection, i.e., $dN_{\pm} = 0$. 
	
	In this regime, the nonlinear damping is not strong and the relaxation to the
	desynchronized state is relatively slow. Therefore, the subsequent photon
	detection typically occurs before the convergence to the desynchronized state
	and $\abs{S_{12}}$ remains always positive.
	Figure~\ref{fig_2}(b) shows that $\theta_{12}$ takes either $\theta_{12} = 0$ or
	$\theta_{12} = \pi$.
	This indicates that the two oscillators immediately attain in-phase coherence
	after the photon detection at P or anti-phase coherence after the photon
	detection at M.
	The negativity and purity are shown in Fig.~\ref{fig_3}(c) and \ref{fig_3}(d),
	respectively, 
	where the negativity is always zero and the purity takes small values between
	$0.03$ and $0.05$,
	indicating that the system is separable and mixed.

	The phase coherence of the two oscillators 
	can also be captured by using the Hushimi 
	Q distribution of the phase difference 
	$\theta = \phi_2 - \phi_1$~\cite{husimi1940some} between the two oscillators,
	Q($\theta$),
	calculated by introducing the two-mode Q
	distribution~\cite{carmichael2007statistical}
	$Q\left(\alpha_{1}, \alpha^*_{1}, 
	\alpha_{2}, \alpha^*_{2}\right)
	=\frac{1}{\pi^{2}}  
	\left\langle 
	\alpha_1, \alpha_2 | \rho | \alpha_1, \alpha_2 \right\rangle$
	with $R_{j}e^{i \phi_j}= \alpha_{j}~(j=1,2)$
	and integrating over $R_1$, $R_2$, and $\phi_1 + \phi_2$.
	Figures~\ref{fig_2}(e) and~\ref{fig_2}(f) show $Q(\theta)$ of the system states
	immediately after the first photon detection at the detectors P and M,
	respectively.
	The peak of $Q(\theta)$ occurs at $\theta=0$ in Fig.~\ref{fig_2}(e) 
	and at $\theta=\pi$ in Fig.~\ref{fig_2}(f), clearly indicating that in-phase and
	anti-phase
	coherence of the two oscillators are induced by the conditional photon
	detection.
	
	\section{Strong quantum regime}
	\begin{figure} [!t]
		\begin{center}
			\includegraphics[width=1\hsize,clip]{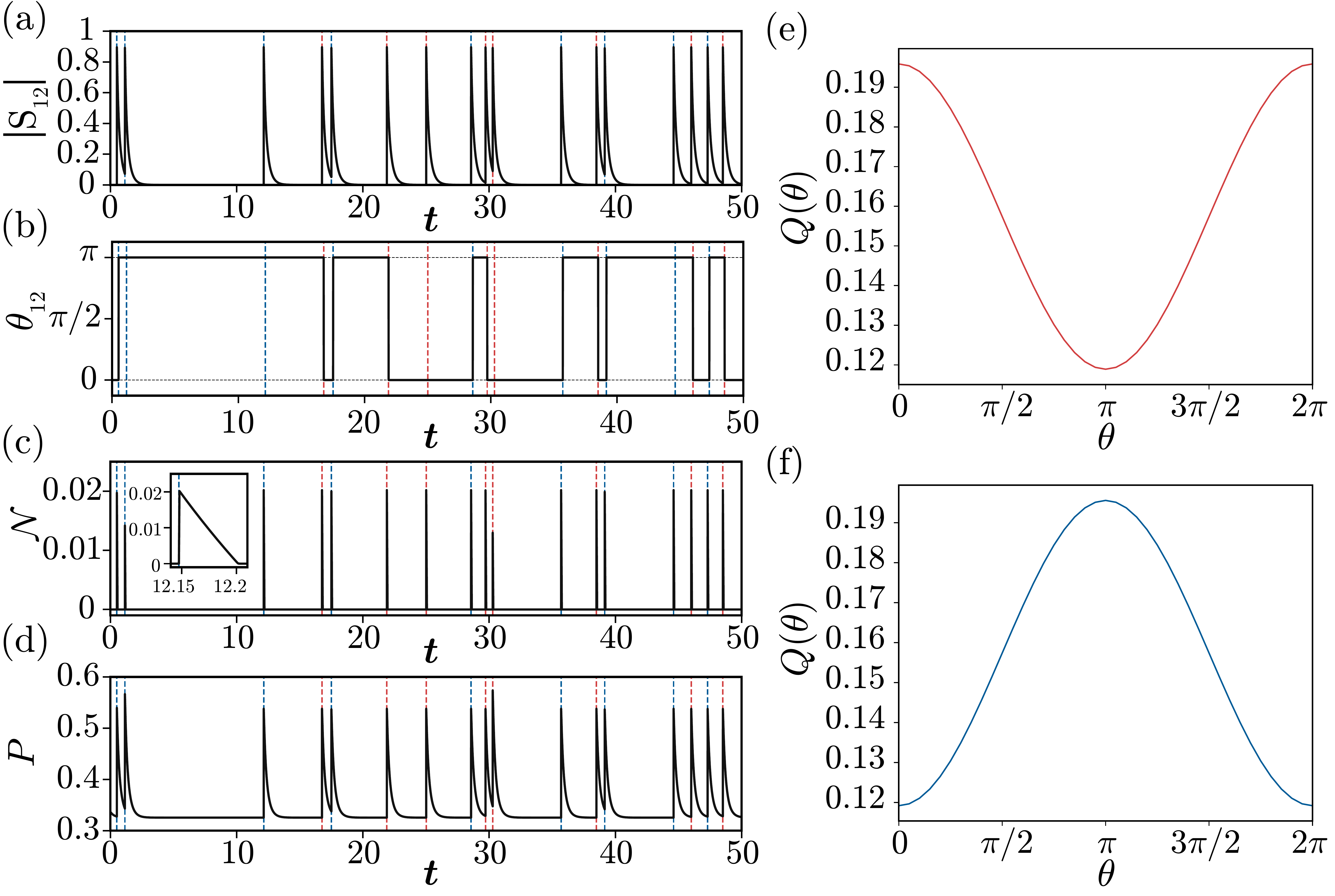}
			\caption{
				Results in the strong quantum regime. The parameters are $(\omega, \gamma_2,
				\gamma_3)/\gamma_1 = (0.5, 50, 0.5)$ with $\gamma_1 = 1$.
				(a-d): Time evolution of 
				(a) absolute value of the normalized correlator $\abs{S_{12}}$,
				(b) averaged phase value $\theta_{12}$,
				(c) negativity $\mathcal{N}$,
				and
				(d) purity $P$.
				(e,f): Q distributions $Q(\theta)$ immediately after the first photon
				detection at (e) P ($t=16.8$) 
				and (f) M ($t=0.52$).
				The photon detection at detectors P and M is indicated by the red- and
				blue-dashed lines in (a-d), respectively.
			}
			\label{fig_3}
		\end{center}
	\end{figure}
	
	We next analyze the quantum SME (\ref{eq:qvdp_sme}) in a stronger quantum
	regime. 
	Figures~\ref{fig_3}(a), \ref{fig_3}(b), \ref{fig_3}(c), and~\ref{fig_3}(d)
	show the evolution of $\abs{S_{12}}$, $\theta_{12}$, $\mathcal{N}$, and $P$,
	respectively.
	As shown in Fig.~\ref{fig_3}(a), $\abs{S_{12}}$ takes large values close to $1$
	immediately after the photon detection, indicating that instantaneous phase
	coherence also arises in this case. 
	In this regime, the nonlinear damping is strong and the system quickly converges
	to the desynchronized steady state of the SME (\ref{eq:qvdp_sme}) when the
	detection does not occur, i.e., $dN_{\pm} = 0$.
	Therefore, the phase coherence quickly disappears and $\abs{S_{12}}$ remains
	zero until the next photon detection occurs.
	
	Similar to Fig.~\ref{fig_2}(b),
	Fig.~\ref{fig_3}(b) shows that $\theta_{12}$ takes either $\theta_{12} = 0$ or
	$\theta_{12} = \pi$.
	Thus, the two oscillators become in-phase coherent after the photon detection at
	P and anti-phase coherent after the photon detection at M.
	Remarkably, Figs.~\ref{fig_3}(c) and \ref{fig_3}(d) show
	that non-zero negativity and purity with values between 
	$0.5$ and $0.6$ are attained instantaneously after the photon detection, 
	indicating that mixed entangled states are obtained in this case.
	However, the quantum entanglement quickly disappears as shown in the inset in
	Fig.~\ref{fig_3}(c). 
	Here, the apparent linear decay of the negativity is due to the
		large  decay rate and the cutoff at zero of the negativity (see e.g.~\cite{guo2013decoherent} for
		discussion about the decay of the negativity).
	Figures~\ref{fig_3}(e) and~\ref{fig_3}(f) show the Q distributions $Q(\theta)$
	of the system states immediately after the first photon detection at the
	detectors P and M, respectively.
	The Q distributions are peaked at $\theta = 0$ and $\theta = \pi$, clearly
	indicating that in-phase and anti-phase coherence of the two oscillators are
	induced also in this case.
	
	\section{Quantum limit}
	
	From the previous numerical results, it is expected that the maximum quantum
	entanglement is attained in the quantum limit, i.e., $\gamma_2 \to \infty$.
	In this limit, we can map the quantum vdP oscillator to an analytically
	tractable two-level system with basis states $\ket{0}$ and
	$\ket{1}$~\cite{lee2014entanglement}, and transform the SME (\ref{eq:qvdp_sme})
	to
	\begin{align}
		\label{eq:qvdp_sme2d}
		d \rho &= \mathcal{L}_0^q\rho dt
		+\mathcal{G}[L^q_{+}]\rho
		\left( dN_{+} - \gamma_{3} \Tr[ L_{+}^{q\dag} L^q_{+} \rho] dt \right)
		+\mathcal{G}[L^q_{-}]\rho 
		\left( dN_{-} - \gamma_{3} \Tr[ L_{-}^{q\dag} L^q_{-} \rho] dt \right),
		\cr
		\mathcal{L}_0^q \rho
		&=
		\sum_{j=1,2} \left( 
		-i \left[  \omega \sigma_{j}^{+} \sigma_{j}^{-}, \rho \right]
		+\gamma_1 \mathcal{D}[\sigma_j^+]\rho 
		+(2 \gamma_1 + \gamma_3) \mathcal{D}[\sigma_j^-]\rho \right),
		~L_{\pm}^q = \frac{1}{\sqrt{2}}(\sigma_1^- \pm \sigma_2^-),
	\end{align}
	with $\sigma_j^- = \ket{0}\bra{1}_j$ 
	and $\sigma_j^+ = \ket{1}\bra{0}_j$ representing the lowering and raising
	operators 
	of the $j$th system ($j = 1,2$), respectively,
	because 
	the transition 
	$\ket{1} \stackrel{2 \gamma_{1}}{\longrightarrow} \ket{2} 
	\stackrel{2 \gamma_{2}}{\longrightarrow} \ket{0}$ can be 
	regarded as 
	$\ket{1} \stackrel{2 \gamma_{1}}{\longrightarrow} \ket{0}$
	when $\gamma_2 \to \infty$.

	The steady state of Eq.~(\ref{eq:qvdp_sme2d}) without detection, i.e., $dN_{\pm}
	= 0$,
	can be analytically obtained, 
	which is given by a diagonal matrix $\rho^{pre} =
	\operatorname{diag}\left(\rho^{pre}_0, \rho^{pre}_1, \rho^{pre}_1, \rho^{pre}_2
	\right)$
	with
	\begin{align}
		\label{eq:pre}
		\rho^{pre}_0 &= \frac{ (k-3)\sqrt{k^{2}+2 k+9} + k^{2} - 2k + 9}{2 k^{2}},
		\cr
		\rho^{pre}_1 &= \frac{3 \sqrt{k^{2}+2 k+9}-k-9}{2 k^{2}},
		\cr
		\rho^{pre}_2 &= \frac{ -(k+3)\sqrt{k^{2}+2 k+9} + k^{2} + 4k + 9}{2 k^{2}}.
	\end{align}
	Note that only a single parameter 
	$k = \gamma_3/\gamma_1$
	specifies the elements of the matrix,
	where we assume $k > 0$, namely, the photon detection occurs with a non-zero
	probability.

	The states $\rho^{pos}_{\pm} = L_{\pm}^q \rho^{pre} L_{\pm}^{q\dag} /
	\Tr[L_{\pm}^q \rho^{pre} L_{\pm}^{q\dag}]$ immediately after the photon
	detection occurs at the detector P ($\rho^{pos}_{+}$) and M ($\rho^{pos}_{-}$)
	can be represented by a density matrix 
	\begin{align}
		\label{eq:pos}
		\rho^{pos}_{\pm} = \rho^{pos}_0 \ket{00}\bra{00} +  
		\rho^{pos}_1 \left(\frac{\ket{01} \pm \ket{10}}{\sqrt{2}}
		\right)\left(\frac{\bra{01} \pm \bra{10}}{\sqrt{2}}\right)
	\end{align}
	with
	\begin{align}
		\label{eq:pos2}
		\rho^{pos}_{0} &= \frac{-3 \sqrt{k^{2}+2 k+9}+k+9}{k\left(\sqrt{k^{2}+2
				k+9}-k-3\right)},
		\cr
		\rho^{pos}_{1} &= \frac{ (k+3)\sqrt{k^{2}+2 k+9}-k^{2}-4 k-9}{
			k\left(\sqrt{k^{2}+2 k+9}-k-3\right)}.
	\end{align}
	Using this result, we can explicitly calculate the normalized correlator
	$S_{12}$ and the Q distribution of the phase difference between the two
	oscillators.
	 If the subsequent photon detection does not occur, the state after
		the
		photon detection in Eq.~(\ref{eq:pos2}) converges to the 
		steady state $\rho_{pre}$
		in Eq.~(\ref{eq:pre}) with the approximate decay rate determined by 
		$\gamma_1$ and $\gamma_3$ in  Eq.~(\ref{eq:qvdp_sme2d}).  In this case, the correlator $S_{12}$ of the states $\rho^{pos}_{\pm}$
	immediately after the photon detection always takes $S_{12} = \pm 1$
	irrespective of the value of $k$ (and then quickly decays). 
	The Q distribution for $\rho^{pos}_{\pm}$ can also be calculated as (similar
	calculation for the Wigner distribution of the phase difference has been
	performed in \cite{lee2014entanglement})
	\begin{align}
		\label{eq:q}
		Q(\theta)[\rho^{pos}_{\pm}] = \frac{1}{2\pi} \pm \frac{ \rho^{pos}_{1} \cos
			\theta}{8}.
	\end{align}
	These results qualitatively agree with the corresponding results in the strong
	quantum regime  in Fig.~\ref{fig_3}.
	It is notable that the dependence of the phase coherence on $k$ can be captured
	by the peak height of 
	$Q(\theta)$
	but not by the normalized correlator $S_{12}$ in the quantum limit.
	Indeed, the element $\rho^{pos}_0 \ket{00}\bra{00}$ in Eq.~(\ref{eq:pos})
	affects $Q(\theta)$ (through $\rho^{pos}_1 = 1 - \rho^{pos}_0$) in
	Eq.~(\ref{eq:q}), whereas it does not affect the value of $S_{12}$.

	The above result indicates that the degree of phase coherence is better
	quantified by the peak height of $Q(\theta)$ rather than 
	$S_{12}$ in strong quantum regimes. 
	This is because $S_{12}$ is defined as a quantum analog of the order parameter
	for the coherence of classical noisy oscillators, 
	which is quantitatively correct only in the semiclassical regime.
	This observation is also important in interpreting the results in the weak and
	strong quantum regimes shown in Figs.~\ref{fig_2} and~\ref{fig_3}, where
	$Q(\theta)$ 
	in the weak quantum regime (Fig.~\ref{fig_2}) are more sharply peaked than those
	in the strong quantum regime (Fig.~\ref{fig_3}),
	while 
	$\abs{S_{12}}$ in Fig.~\ref{fig_2} takes smaller values than that in
	Fig.~\ref{fig_3}.
	Thus, $S_{12}$ may not work well for comparing phase coherence between different
	quantum regimes.
	
	In the quantum limit,
	the symmetric superpositions $\ket{S} = (\ket{01} + \ket{10})/\sqrt{2}$
	and $\ket{A} = (\ket{01} - \ket{10})/\sqrt{2}$ can be regarded as the
	in-phase and anti-phase synchronized states, because the corresponding
	distributions 
	$Q(\theta)[\ket{S}\bra{S}] = \frac{1}{2\pi} + \frac{\cos \theta}{8}$ and 
	$Q(\theta)[\ket{A}\bra{A}] =\frac{1}{2\pi} - \frac{\cos \theta}{8}$
	are peaked at $\theta = 0$ and $\theta = \pi$, respectively.
	As $\ket{A}$ and $\ket{S}$ are 
	dark states with respect to 
	$L^{q}_{+}$ and $L^{q}_{-}$, i.e.,  
	$L^{q}_{+} \ket{A} = 0$ and $L^{q}_{-} \ket{S} = 0$,
	the photon detection at the detector P annihilates the 
	anti-phase-synchronized state $\ket{A}$ and 
	creates the in-phase synchronized state $\ket{S}$ with 
	$S_{12} = 1~(\theta_{12} = 0)$,
	while the photon detection at the detector M
	annihilates 
	$\ket{S}$ and creates $\ket{A}$ with 
	$S_{12} = -1~(\theta_{12} = \pi)$.
	
	\begin{figure} [!t]
		\begin{center}
			\includegraphics[width=1\hsize,clip]{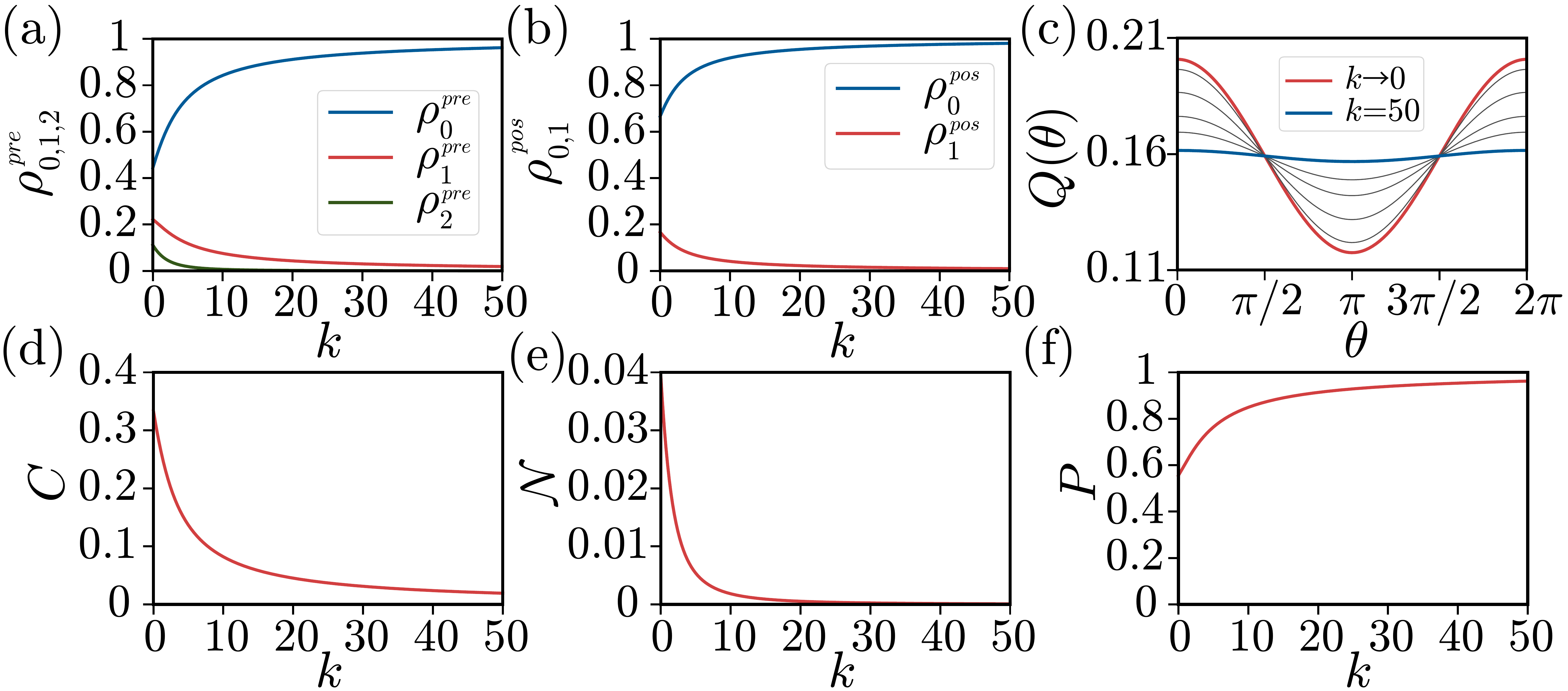}
			\caption
			{
				Dependence of the results on the parameter 
				$k = \gamma_3/\gamma_1$ of the two-level system
				in the quantum limit $\gamma_2 \to \infty$.
				(a) Elements of $\rho^{pre}$.
				(b) Elements of $\rho^{pos}$.
				(c) $Q(\theta)[\rho^{pos}_{+}]$ distributions  
				of the phase difference of two oscillators
				for $k \to 0$~(red line), 
				$k = 0.5, 2, 5, 10$~(gray lines from the red line to the blue line), and
				$k=50$~(blue line) are shown.
				(d) Concurrence.
				(e) Negativity.
				(f) Purity.
			}
			\label{fig_4}
		\end{center}
	\end{figure}

	Figures~\ref{fig_4}(a),~\ref{fig_4}(b), and~\ref{fig_4}(c) 
	show the dependence of the elements 
	$\rho^{pre}$ and $\rho^{pos}$ and $Q(\theta)[\rho^{pos}_{+}]$
	on $k$, respectively (we only plot $Q(\theta)[\rho^{pos}_{+}]$ because
	$Q(\theta)[\rho^{pos}_{-}] = Q(\theta + \pi)[\rho^{pos}_{+}]$).
	As shown in Fig.~\ref{fig_4}(a), $\rho^{pre}_1$ and $\rho^{pre}_2$ take larger
	values when $k$ is smaller.
	When $k \to 0$, $\rho^{pre}_1$ and $\rho^{pre}_2$ approach the supremum values,
	$\rho^{pre}_{1} \to \frac{2}{9}$ and $\rho^{pre}_{2} \to
	\frac{1}{9}~(\rho^{pre}_{0} \to \frac{4}{9})$, 
	corresponding to the completely incoherent
	steady state of the two decoupled quantum vdP oscillators in the quantum limit,
	i.e., 
	$ \rho^{pre} \to (\frac{2}{3}\ket{0}\bra{0} + \frac{1}{3}\ket{1}\bra{1}) \otimes
	(\frac{2}{3}\ket{0}\bra{0} + \frac{1}{3}\ket{1}\bra{1})$, where $Q(\theta)$ is
	uniform~\cite{lee2013quantum,lee2014entanglement}.
	Therefore,
	$\rho^{pos}_1$ approaches the supremum value,
	$\rho^{pos}_1 \to \frac{1}{3}~(\rho^{pos}_0 \to \frac{2}{3})$,
	as shown in Fig.~\ref{fig_4}(b), and
	$Q(\theta)[\rho^{pos}_{+}]$ 
	exhibits the maximum peak
	as shown in Fig.~\ref{fig_4}(c),
	indicating that the maximum phase coherence is obtained. 
	In the opposite limit, $ k \to \infty$,
	$\rho^{pre}$ converges to the two-mode vacuum state
	$\rho^{pre} \to \ket{00}\bra{00}$, i.e.,
	$\rho^{pre}_1, \rho^{pre}_2 \to 0~(\rho^{pre}_0 \to 1)$,
	resulting in 
	$\rho^{pos}_1 \to 0~(\rho^{pos}_0 \to 1)$ and
	the uniform distribution
	$Q(\theta)[\rho^{pos}_{+}] \to \frac{\pi}{2}$.
	Note that we can only consider the limit $k \to 0$
	(no photon detection occurs when $k=0$), and that the two-mode vacuum state in the $k \to \infty$ limit is not a limit cycle.
	
	In addition to the negativity ${\mathcal N}$ and purity $P$, the quantum
	entanglement of the density matrix $\rho_{pos}$ in Eq.~(\ref{eq:pos})
	can also be quantified using the concurrence~\cite{wootters1998entanglement},
	$C = \max \left(0, \lambda_{1}-\lambda_{2}-\lambda_{3}-\lambda_{4}\right)$,
	where $\lambda_{1}, \lambda_{2}, \lambda_{3}$, and $\lambda_{4}$
	are the square roots of the eigenvalues of $\rho \tilde{\rho}$ 
	with
	$\tilde{\rho}=\left(\sigma_{y} \otimes \sigma_{y}\right)
	\rho^{*}\left(\sigma_{y} \otimes \sigma_{y}\right)$
	in decreasing order.
	The concurrence $C$ takes a non-zero value when the two oscillators are
	entangled with each other ($C \in[0,1]$ by definition).
	
	Figures~\ref{fig_4}(d),~\ref{fig_4}(e), and~\ref{fig_4}(f) show the dependence
	of $C$, $\mathcal{N}$ and $P$ on $k$
	for $\rho_{\pm}^{pos}$, respectively.
	Note that $C$, $\mathcal{N}$, and $P$ take the same values for both $\rho_{+}$
	and $\rho_{-}$.
	In the limit $k \to 0$, $C$, $\mathcal{N}$, and $P$ approach the 
	upper bounds as
	$C \to \frac{1}{3}$, $\mathcal{N} \to \frac{\sqrt{5}-2}{6}$,
	and $P \to \frac{5}{9}$. 
	In the opposite limit $k \to \infty$, 
	these values converge as $C \to 0$, $\mathcal{N} \to 0$, and $P \to 1$, which
	corresponds to those quantities for the two-mode vacuum states.

	Photon detection occurs less frequently when $k$ is smaller, because the
	probability of the photon detection in the interval $dt$ at detectors P or M is
	given by 
	$k \gamma_1 \operatorname{Tr}\left[L_{ \pm}^{\dagger} L_{ \pm} \rho
		\right]dt$.
	Therefore, on average, infinitely-long observation time is required before the
	photon detection to approach
	the upper bounds for the degree of phase coherence and quantum entanglement in
	the limit $k \to 0$.
	
	%

		\section{Concluding remarks}
	We have analyzed two decoupled quantum van der Pol oscillators
	and demonstrated that quantum measurement backaction of conditional photon
	detection 
	induces instantaneous phase synchronization of the oscillators.
	In-phase or anti-phase coherence between the oscillators has been observed
	instantaneously after the photon detection, which decays gradually in the weak
	quantum regime or quickly in the strong quantum regime
	until the next photon detection.
	In the strong quantum regime, short-time increase in the quantum entanglement
	has also been observed.
	In the quantum limit, we analytically obtained the upper bounds for the
	increase in the quantum entanglement and phase coherence.
	
	In this paper, we presented only the results for the case with two
		identical oscillators under the measurement without inefficiency. 
		For the case with oscillators whose natural frequencies are slightly different,
		we also confirmed numerically that almost the same results as in
		Figs.~\ref{fig_2} and~\ref{fig_3} are obtained except for the phase difference
		shown in Fig.~\ref{fig_2}(b) or~\ref{fig_3}(b); in this case, the phase
		difference between the oscillators increases or decreases with a constant rate
		(i.e., with the frequency difference) between the jumps to $0$ or $\pi$ caused
		by the photon detection.
		%
		%
		The results in the quantum limit analyzed in Sec.~V are also independent of the
		natural frequencies of the two oscillators.
		As for the measurement inefficiency \cite{jacobs2006straightforward}, we
		confirmed numerically that it mainly affects the frequency of photon detection.
		This is because the probability of photon detection is proportional to the
		measurement efficiency, whereas the system state just after the photon detection
		is not strongly affected as the $dN_{\pm}$ terms in Eq.(\ref{eq:qvdp_sme}) are
		independent of the measurement efficiency.  
	
	Recently, physical implementations of the quantum vdP oscillator with ion trap
	systems~\cite{lee2013quantum,lee2014entanglement} and optomechanical
	systems~\cite{walter2014quantum, walter2015quantum} have been discussed.
	The additional single-photon dissipation
	and photon detectors can also be introduced~\cite{carmichael2007statistical,
		wiseman2009quantum}.
	The physical setup considered in the present study does not require explicit
	mutual coupling between the oscillators. Therefore, it can, in principle, be
	implemented by using existing experimental methods and provide a method for
	generating phase-coherent states of quantum limit-cycle oscillators.

	\acknowledgments{
		The numerical simulations are performed by using the QuTiP numerical
		toolbox \cite{johansson2012qutip,*johansson2013qutip}. 
		We acknowledge JSPS KAKENHI JP17H03279, JP18H03287, JPJSBP120202201, JP20J13778,
		and JST CREST JP-MJCR1913 for financial support. }
	
	\appendix
	
	\section{SLH framework}

	In this Appendix, we derive the SME~(\ref{eq:qvdp_sme})
	using the SLH framework to describe cascade and concatenate connections
	of the quantum system components \cite{gough2009series,combes2017slh}.
	In this framework, the parameters in the time evolution 
	of a quantum system $\rho$ are specified by $\bm{G} = (\bm{S}, \bm{L}, H)$
	with
	\begin{align}
		\bm{S} =
		\begin{pmatrix} 
			S_{11} \cdots S_{1n} \\
			\vdots \hspace{5mm} \vdots \hspace{5mm} \vdots \\
			S_{n1} \cdots S_{nn}
		\end{pmatrix},~
		\bm{L} = \begin{pmatrix} L_{1} \\ \vdots \\ L_{n}\end{pmatrix},
	\end{align}
	where $\bm S$ is the scattering matrix with operator entries 
	satisfying $\bm{S}^{\dag} \bm{S} = \bm{S} \bm{S}^{\dag} = \bm{I}^n$,
	$\bm{L}$ is a coupling vector with operator entries,
	and $H$ is a self-adjoint operator referred to as the system Hamiltonian.
	We denote by $\bm{I}^n$ an identity matrix with $n$ dimensions.

	\begin{figure} [!t]
		\begin{center}
			\includegraphics[width=0.8\hsize,clip]{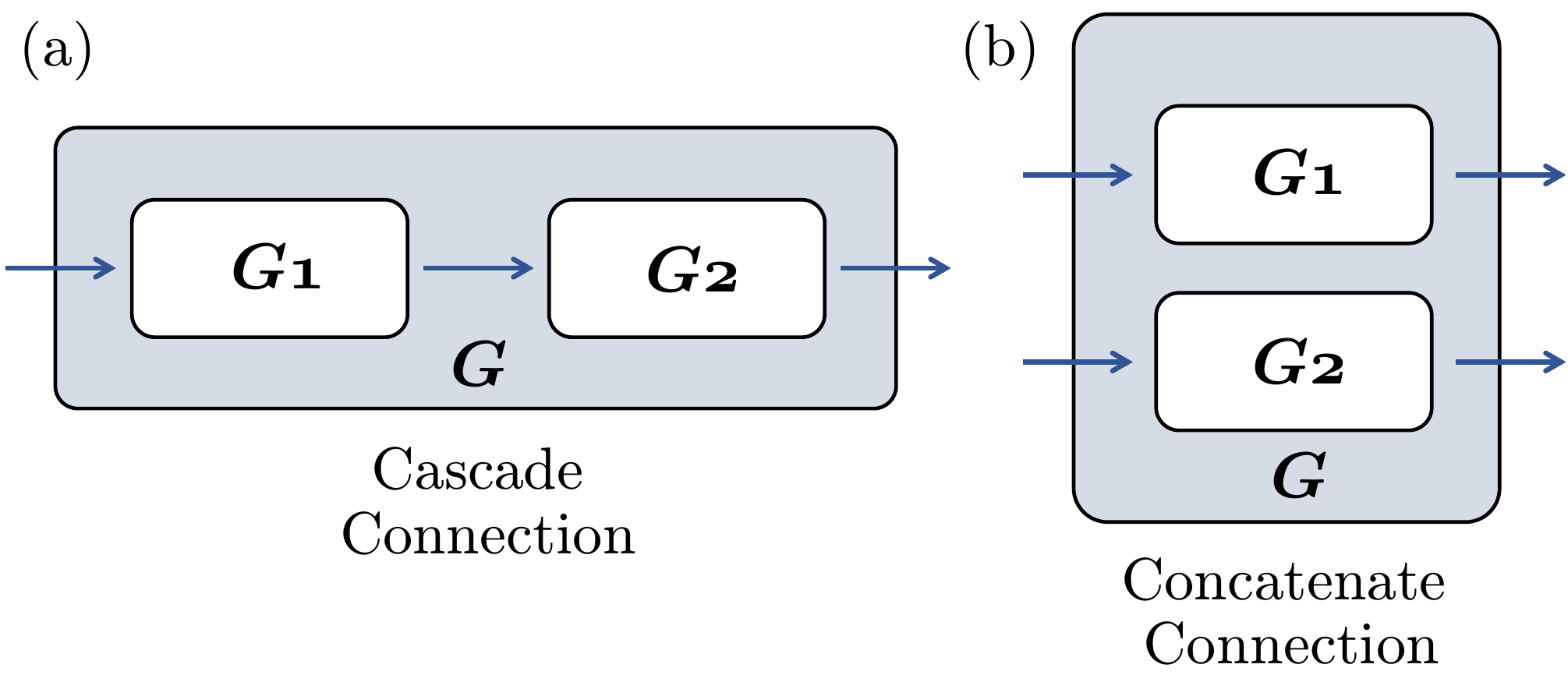}
			\caption{
				(a) Cascade and concatenate connection
				of the two system components $G_1$ and $G_2$.	
			}
			\label{fig_5}
		\end{center}
	\end{figure}

	With these parameters, the time evolution of the system 
	obeys the master equation
	\begin{align}
		\label{eq:me_slh}
		\frac{d\rho}{dt} = - i [H, \rho] + \sum_{i=1}^{n} \mathcal{D}[L_i]\rho,
	\end{align}
	where $\bm{S}$ is involved in 
	the calculation of the cascade and concatenation products and
	has an important role in determining the forms of $H$ and $\bm{L}$
	of the whole network system consisting of the system components.
	This specification of parameters is based on Hudson-Parthasarathy's
	work~\cite{hudson1984quantum}.
	
	The cascade product (Fig.~\ref{fig_5}(a)) of 
	$\bm{G}_1 = (\bm{S}_1, \bm{L}_1, H_1)$ 
	and $\bm{G}_2 = (\bm{S}_2, \bm{L}_1, H_2)$ is given by
	\begin{align}
		\bm{G}_{1} \triangleleft \bm{G}_{2} = 
		\left( \bm{S}_2 \bm{S}_1 , \bm{L}_2 + \bm{S}_2 \bm{L}_{1}, H_1 + H_2
		+\frac{1}{2i}\left(    	
		\bm{L}_2^{\dag} \bm{S}_{2} \bm{L}_{1} - \bm{L}_{1}^{\dag} \bm{S}_{2}^{\dag}
		\bm{L}_{2}  
		\right)\right),
	\end{align}
	and the concatenation product (See Fig.~\ref{fig_5}(b)) of $\bm{G}_1$ and
	$\bm{G}_2$ is given by 
	\begin{align}
		\bm{G}_{1} \boxplus \bm{G}_{2} = \left( \begin{pmatrix} \bm{S}_{1} &0 \\ 0 &
			\bm{S}_{2} \end{pmatrix},			
		\begin{pmatrix} \bm{L}_{1} \\ \bm{L}_{2} \end{pmatrix}, H_{1} + H_{2} \right).
	\end{align}

	Our aim is to derive the SME~(\ref{eq:qvdp_sme})
	of the physical setup depicted in Fig.~1  
	\cite{gough2009series,combes2017slh}. 
	To this end, we denote $\bm{G}^{QVDP}_j$ as the parameters of the $j$th quantum
	vdP oscillator with 
	an additional single-photon dissipation,%
	\begin{align}
		\bm{G}^{QVDP}_j =  
		\left( \bm{I}^3, 
		\begin{pmatrix} \sqrt{\gamma_{1}} a_{j}^{\dag} \\ \sqrt{\gamma_{2}} a_{j}^{2} \\
			\sqrt{\gamma_{3}} a_{j} \end{pmatrix},
		\omega a_{j}^{\dag}a_{j} \right).
	\end{align}

	The concatenate connection
	of $\bm{G}^{QVDP}_1$ and $\bm{G}^{QVDP}_2$ is
	\begin{align}
		\bm{G}^{QVDP}_{1} \boxplus \bm{G}^{QVDP}_{2} =  
		\left( \bm{I}^6, 
		\begin{pmatrix} 
			\sqrt{\gamma_{1}} a_{1}^{\dag} \\ \sqrt{\gamma_{1}} a_{2}^{\dag} \\
			\sqrt{\gamma_{2}} a_{1}^{2} \\ \sqrt{\gamma_{2}} a_{2}^{2} \\   
			\sqrt{\gamma_{3}} a_{1} \\ \sqrt{\gamma_{3}} a_{2}  
		\end{pmatrix},
		\sum_{j=1,2} \omega a_{j}^{\dag}a_{j} \right),
	\end{align}
	where we have changed the order of the elements in $\bm{L}$ 
	for simplicity of notation.
	
	In this study, we consider a 50:50 beam splitter.
	The parameters of the beam splitter $\bm{G}^{BS}$
	for the output fields of the two 
	baths
	\begin{align}
		\bm{G}^{BS} =  
		\left( \begin{pmatrix} 
			\bm{I}^4 & \bm{O}^{42} \\
			\bm{O}^{24} & 
			\begin{pmatrix} 
				\frac{1}{\sqrt{2}} & -\frac{1}{\sqrt{2}} \\ 
				\frac{1}{\sqrt{2}} & \frac{1}{\sqrt{2}}
			\end{pmatrix}
		\end{pmatrix}, 
		0,
		0 \right),
	\end{align}
	where we denote by $\bm{O}^{nm}$ a zero matrix with the dimensions $n \times m$.
	
	The cascading connection of the two above-mentioned components 
	is given by
	\begin{align}
		\label{eq:Gsys}
		\bm{G}^{QVDP}_{1} \boxplus \bm{G}^{QVDP}_{2} \triangleleft \bm{G}^{BS}
		=
		\left( 
		\begin{pmatrix} 
			\bm{I}^4 & \bm{O}^{42} \\
			\bm{O}^{24} & 
			\begin{pmatrix} 
				\frac{1}{\sqrt{2}} & -\frac{1}{\sqrt{2}} \\ 
				\frac{1}{\sqrt{2}} & \frac{1}{\sqrt{2}}
			\end{pmatrix}
		\end{pmatrix}, 
		\begin{pmatrix} 
			\sqrt{\gamma_{1}} a_{1}^{\dag} \\ \sqrt{\gamma_{1}} a_{2}^{\dag} \\
			\sqrt{\gamma_{2}} a_{1}^{2} \\ \sqrt{\gamma_{2}} a_{2}^{2} \\   
			\sqrt{\gamma_{3}} \frac{a_{1} - a_{2}}{\sqrt{2}}  \\ \sqrt{\gamma_{3}}
			\frac{a_{1} + a_{2}}{\sqrt{2}}  
		\end{pmatrix},
		\sum_{j=1,2} \omega a_{j}^{\dag}a_{j} \right).
	\end{align}
	
	Using transformation 
	$
	\mathcal{D}[\frac{a_{1} + a_{2}}{\sqrt{2}} ]\rho + \mathcal{D}[\frac{a_{1} -
		a_{2}}{\sqrt{2}}]\rho
	= \mathcal{D}[a_1]\rho + \mathcal{D}[a_2]\rho  
	$,
	the quantum master equation 
	(\ref{eq:me_slh}) with 
	the parameters given in Eq.~(\ref{eq:Gsys})
	gives  $ d\rho = \mathcal{L}_0\rho dt$
	of the SME~(\ref{eq:qvdp_sme}).
	Then, using the quantum filtering theory
	\cite{van2005feedback,bouten2007introduction},
	SME~(\ref{eq:qvdp_sme}) can be obtained.
	%
	%
	%
	
\end{document}